\documentclass[twocolumn, astrosymb]{aastex701}

\usepackage{textcomp}
\usepackage{amssymb}
\usepackage{amsmath}
\usepackage{comment}
\usepackage{wasysym}
\usepackage{CJK}
\usepackage{verbatim}
\usepackage{enumerate}
\usepackage{booktabs}
\usepackage{graphicx}

\begin{document}
\begin{CJK*}{UTF8}{gbsn}

\title{Episodic Feedback in Triple AGN Candidate SDSS J0849+1114 Revealed by Extended ionized gas}

\correspondingauthor{Xiaoyu Xu, Meicun Hou, Zhiyuan Li}
\email{xuxy95@nju.edu.cn, houmc@nju.edu.cn, lizy@nju.edu.cn}

\author[0000-0003-0970-535X]{Xiaoyu Xu (许啸宇)}
\affiliation{School of Astronomy and Space Science, Nanjing University, Nanjing 210023, China}
\affiliation{Key Laboratory of Modern Astronomy and Astrophysics, Nanjing University, Nanjing 210023, China}
\email{xuxy95@nju.edu.cn}  

\author[0000-0001-9062-8309]{Meicun Hou}
\affiliation{Institute of Science and Technology for Deep Space Exploration, Suzhou Campus, Nanjing University, Suzhou 215163, China}
\email{houmc@nju.edu.cn}  

\author[0000-0003-0355-6437]{Zhiyuan Li}
\affiliation{School of Astronomy and Space Science, Nanjing University, Nanjing 210023, China}
\affiliation{Key Laboratory of Modern Astronomy and Astrophysics, Nanjing University, Nanjing 210023, China}
\email{lizy@nju.edu.cn}

\author[0000-0001-8492-892X]{Sijia Peng}
\affiliation{Shanghai Astronomical Observatory, Chinese Academy of Sciences, Shanghai 200030, China}
\email{sjpeng@shao.ac.cn}

\author[0000-0002-6738-3259]{Zhao Su}
\affiliation{School of Astronomy and Space Science, Nanjing University, Nanjing 210023, China}
\affiliation{Key Laboratory of Modern Astronomy and Astrophysics, Nanjing University, Nanjing 210023, China}
\email{suzhao@smail.nju.edu.cn}

\author[0000-0002-7172-6306]{Zongnan Li}
\affiliation{National Astronomical Observatory of Japan, 2-21-1 Osawa, Mitaka, Tokyo, 181-8588, Japan}
\affiliation{East Asian Core Observatories Association (EACOA) Fellow}
\email{zongnan.li@astro.nao.ac.jp}

\author[0000-0002-1620-0897]{Fuyan Bian}
\affiliation{European Southern Observatory, Alonso de Cordova 3107, Casilla 19001, Vitacura, Santiago 19, Chile}
\email{fbian@eso.org}

\author[0000-0003-4874-0369]{Junfeng Wang}
\affiliation{Department of Astronomy, Xiamen University, Xiamen, Fujian 361005, China}
\email{jfwang@xmu.edu.cn}

\begin{abstract}

Galaxy mergers funnel gas toward the nuclei, igniting starbursts and active galactic nuclei (AGNs). 
The AGN feedback can reshape the host galaxy and regulate both star formation and super-massive black-hole (SMBH) accretion.
Using VLT/MUSE integral-field spectroscopy, we conduct a spatially resolved study of the triple-AGN candidate SDSS J0849+1114.
Extended ionized gas structures ($>10$ kpc from nucleus A) primarily associated with tidal tails are detected.
Meanwhile, two distinct ionized gas outflows are revealed. 
One extends over $>5$ kpc around nuclei A with a kinetic power of $\dot{E}_{\rm out,A} = 3.0\times10^{42}\rm\, erg\, s^{-1}$, which might be driven by the radio jet. 
The other outflow extends $\sim 5.9$ kpc around nucleus C, with a kinetic power of $\dot{E}_{\rm out,C} = 2.0\times10^{40}\rm\, erg\, s^{-1}$. 
High [O~{\sc{iii}}]/H$\alpha$ and [N~{\sc{ii}}]/H$\alpha$ ratios in the tidal gas require that nucleus A radiated at a high accretion rate with $L_{\rm A,bol} \sim 0.1$--$0.5\,L_{\rm Edd,A}$ at least $\sim3$--$\times10^{4}\rm\,yr$ ago, $20$--$100$ times brighter than today.
Combined with multi-wavelength constraints, we find evidence for episodic AGN feedback that expelled circumnuclear gas and rapidly quenched accretion. 
This triple AGN candidate demonstrates how AGN feedback can self-regulate black hole growth and impact hosts during mergers.

\end{abstract}

\keywords{Interacting galaxies(802) --- Galaxy winds (626) --- Interstellar medium (847)}

\section{Introduction} \label{sec:intro}

Galaxy mergers drive gas into galactic centers, igniting starbursts and feeding the central super-massive black hole (SMBH), thus activating an active galactic nucleus (AGN) \citep[][]{1988ApJ...325...74S,2005Natur.433..604D}. 
Both star formation (SF) and AGN activity launch kpc-scale winds \citep[e.g.][]{2005ARA&A..43..769V,2008ApJS..175..356H, 2017MNRAS.472L.109A}. 
In nearby galaxies, especially mergers, multiphase SF winds have been observed that extend several kpc and reach velocities of a few $\times100\rm\,km\,s^{-1}$ \citep[e.g.][]{1990ApJS...74..833H,2000ApJS..129..493H,2005ApJS..160..115R,2013ApJ...768...75R}.
Multiphase AGN outflows have now been unambiguously identified over a range of scales and redshifts \citep[e.g.][]{2012ARA&A..50..455F,2021NatAs...5...13L}.

AGN outflows are expected to substantially influence the host galaxy’s growth and evolution through interactions with the interstellar and intergalactic medium \citep[ISM and IGM;][]{1998A&A...331L...1S, 2012ARA&A..50..455F,2015ARA&A..53..115K}. 
In the negative feedback scenario, the heated or expelled cold gas reduces SF by limiting available fuel \citep[e.g.][]{1998A&A...331L...1S,2008ApJS..175..356H,2012ARA&A..50..455F,2012RAA....12..917S}. 
Alternatively, some studies indicate that outflows can compress ambient gas to promote SF, suggesting a positive feedback mechanism \citep[e.g.][]{2013ApJ...772..112S,2017MNRAS.468.4956Z}.

In rare but fascinating instances, more than two AGNs may coexist if the merging process involves multiple galaxies.
Although dozens of triple AGN candidates at low redshift have been reported \citep[e.g.][]{2011ApJ...737..101L}, spatially resolved studies of associated galactic-scale ionized gas and gas outflows remain scarce.

SDSS J084905.51+111447.2 (hereafter J0849+1114), at redshift $z = 0.0775$, features three optical stellar nuclei within a projected radius of $\sim$5 kpc and exhibits a disturbed morphology \citep[][]{2019ApJ...887...90L,2019ApJ...883..167P}. 
Both \cite{2019ApJ...887...90L} and \cite{2019ApJ...883..167P} classified all three optical nuclei as Seyfert 2 AGNs from narrow emission-line diagnostics \citep[BPT diagram][]{1981PASP...93....5B}.
Chandra X-ray observations supported all three nuclei should host AGNs \citep[][]{2019ApJ...887...90L,2019ApJ...883..167P}.
In the radio band, nuclei A and C, along with some extended features associated with them, were detected at 9 GHz using Very Large Array (VLA) observations \citep[][]{2019ApJ...887...90L}.
Subsequently, \cite{2022ApJ...934...89P} detected nuclei A and C at 3.0, 6.0, and 15 GHz for the first time, revealing that both nuclei exhibit double-sided jets, which could well signify merger-driven AGN feedback.
\cite{2019ApJ...883..167P} also found broad [O~{\sc{iii}}]$\lambda\lambda$4959,5007 emission lines in all three nuclei, which is suggestive of ionized gas outflows.

Here, we analyze the extended ionized gas and outflows in J0849+1114 using Very Large Telescope (VLT)/MUSE data. 
The observations capture a fading AGN and feedback signatures that appear to modulate accretion onto the central super-massive black hole.
In Section \ref{sec:data}, we describe the methodology. 
Main results are described in Section \ref{sec:result}. 
Discussions and conclusions are presented in Section \ref{sec:discussion} and Section \ref{sec:summary}, respectively.

A luminosity distance of $355.5\rm\, Mpc$ and a physical scale of $1.48\rm\, kpc/\arcsec$ for J0849+1114 are adopted, assuming $\Omega_{\rm m} = 0.286$, $\Omega_{\rm \Lambda} = 0.714$, and $H_{0} = 69.6\rm\, km\, s^{-1}\, Mpc^{-1}$.

\section{Observations and Data Reduction} \label{sec:data}

\begin{figure*}[ht!]
\includegraphics[width=1\textwidth]{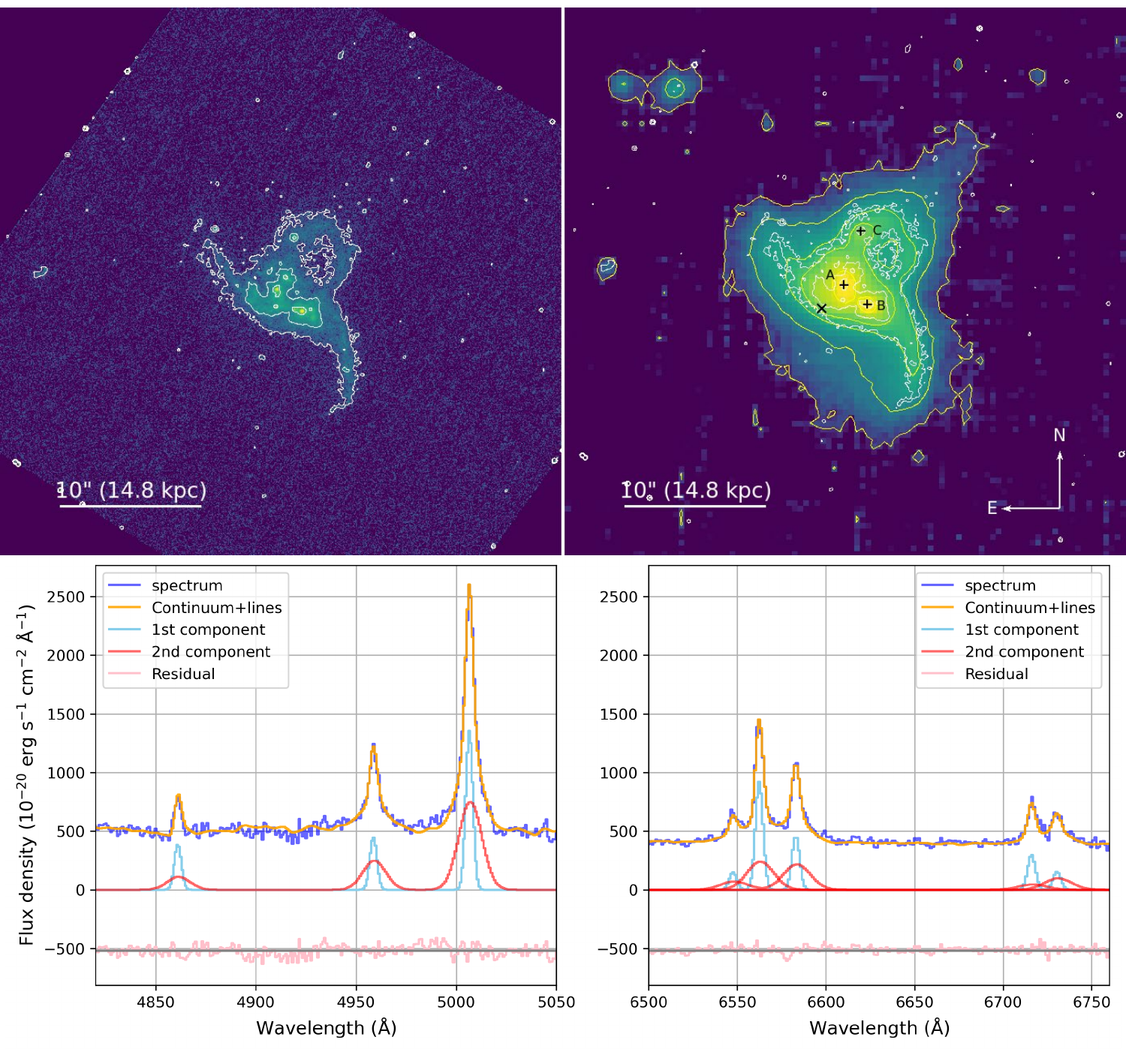}
\caption{Top left: HST/WFC3 UVIS/F336W (U-band) image \citep[][]{2019ApJ...887...90L,2022ApJ...934...89P}.
The white contours are at levels of 1, 5, 15, and 100 $\mu \rm Jy$ arcsec$^{-2}$.
Top right: Continuum image integrated over the rest-frame wavelength range $5500$ to $5800\rm\,\AA$ from MUSE data.
The yellow contours are at levels of 20, 40, 100, 200, and 800 $\times 10^{-20} \rm\,erg\,s^{-1}\,cm^{-2}\,\AA^{-1}$.
White contours are the same as those in the top left panel.
The locations of nuclei A, B, and C are marked with black crosses.
Bottom: Zoomed-in spectra for the rest-frame wavelength ranges $4820$--$5050\rm\,\AA$ (left) and $6500$--$6760\rm\,\AA$ (right), extracted from the spaxel marked with a black ``X” in the top right panel.
The data is shown as a blue line, while the continuum plus emission line model is displayed as an orange line.
The first and second Gaussian components are represented by light blue and red lines, respectively.
For clarity, the residuals (in pink) are shifted below zero.
}
\label{fig:hst_muse_spec}
\end{figure*}

J0849+1114 was observed in 2022 using MUSE in wide field mode \citep[WFM;][]{2010SPIE.7735E..08B,2014Msngr.157...13B} under program 108.21ZY.012 (PI: Bian). 
The spaxel size of MUSE data is 0.\arcsec2, but the spatial resolution is limited by the seeing ($\sim 1\arcsec$).
The average spectral resolution is R $\sim$ 3000 ($\sim 50\rm\, km/s$ for velocity dispersion).
We downloaded the data from ESO archive\footnote{$\rm http://archive.eso.org/eso/eso\_archive\_main.html$} and processed them using EsoReflex \citep{2013A&A...559A..96F} with the MUSE workflow v2.8.5 \citep{2020A&A...641A..28W}.
After inspecting all datasets, we found that only three exposures (IDs: MUSE.2022-01-10T04:33:05.198, MUSE.2022-01-10T04:18:53.472, and MUSE.2022-01-18T02:42:25.369) were suitable for studying the extended ionized gas. 
These three datasets were combined into a final data cube with a total exposure time of 1500 seconds.
The data cube was corrected for Milky Way extinction, using $E_{\rm B-V} = 0.029$ (from NED\footnote{https://ned.ipac.caltech.edu}) and the extinction law of \cite{1999PASP..111...63F}.

To better analyze faint extended structures, we rebinned the data by a factor of 2 (i.e. a spaxel of 0.\arcsec4).
The Penalized Pixel-Fitting method \citep[pPXF;][]{2004PASP..116..138C} was applied to fit the spectrum of each spaxel within the rest-frame wavelength range $\rm 4700$--$6800\,\AA$.
The stellar continuum was modeled using the stellar population templates from \cite{2003MNRAS.344.1000B} (BC03).
After determining the best-fit stellar continuum, two Gaussian components were used to fit the main gas emission lines: H$\beta$, [O~{\sc{iii}}]$\lambda\lambda$4959,5007, [O~{\sc{i}}]$\lambda$6300, H$\alpha$, [N~{\sc{ii}}]$\lambda\lambda$6548,6583, and [S~{\sc{ii}}]$\lambda\lambda$6716,6731.
For each Gaussian component, the velocity ($v$) and velocity dispersion ($\sigma$) were assumed to be the same across all emission lines following \cite{2019ApJ...883..167P,2019ApJ...887...90L}.
However, in some galaxies, different emission lines may exhibit different $v$ and $\sigma$, especially for [O~{\sc{iii}}]$\lambda\lambda$4959,5007 \citep[e.g.][]{2021ApJ...906L...6R,2022ApJ...933..110X}.
We tested this scenario in our data.
In most spaxels, the $v$ and $\sigma$ of the [O~{\sc{iii}}] doublet differ from those of the other lines by less than $50\rm\,km\,s^{-1}$.
Such small offsets indicate that assuming common kinematics introduces no significant bias to our results.
We imposed a lower limit of $\sigma_{1} \geq 50\rm\, km\,s^{-1}$ for the first component and required the second to be broader ($\sigma_{2}>\sigma_{1}$). 
To ensure robust detections, we retained only components with $\rm S/N \geq 5$ and line flux $\geq 1 \times 10^{-18}\rm\, erg\, s^{-1}\, cm^{-2}$ for each spaxel.
An example of the emission line fitting is shown in the bottom panel of Figure~\ref{fig:hst_muse_spec}.

Extinction correction for emission lines was performed using the Balmer decrement (the flux ratio of H$\alpha$ and H$\beta$). 
The attenuation law from \cite{2000ApJ...533..682C} with $\rm R_{V}=3.1$ and an intrinsic Balmer decrement $\rm H\alpha / H\beta=3.1$ for AGNs \citep[][]{2006agna.book.....O} were adopted.

\section{Results} \label{sec:result}

\subsection{Ionized gas distribution and kinematics} \label{subsec: flux and kinematics}

\begin{figure*}[ht!]
\includegraphics[width=1\textwidth,trim=0 30 0 20]{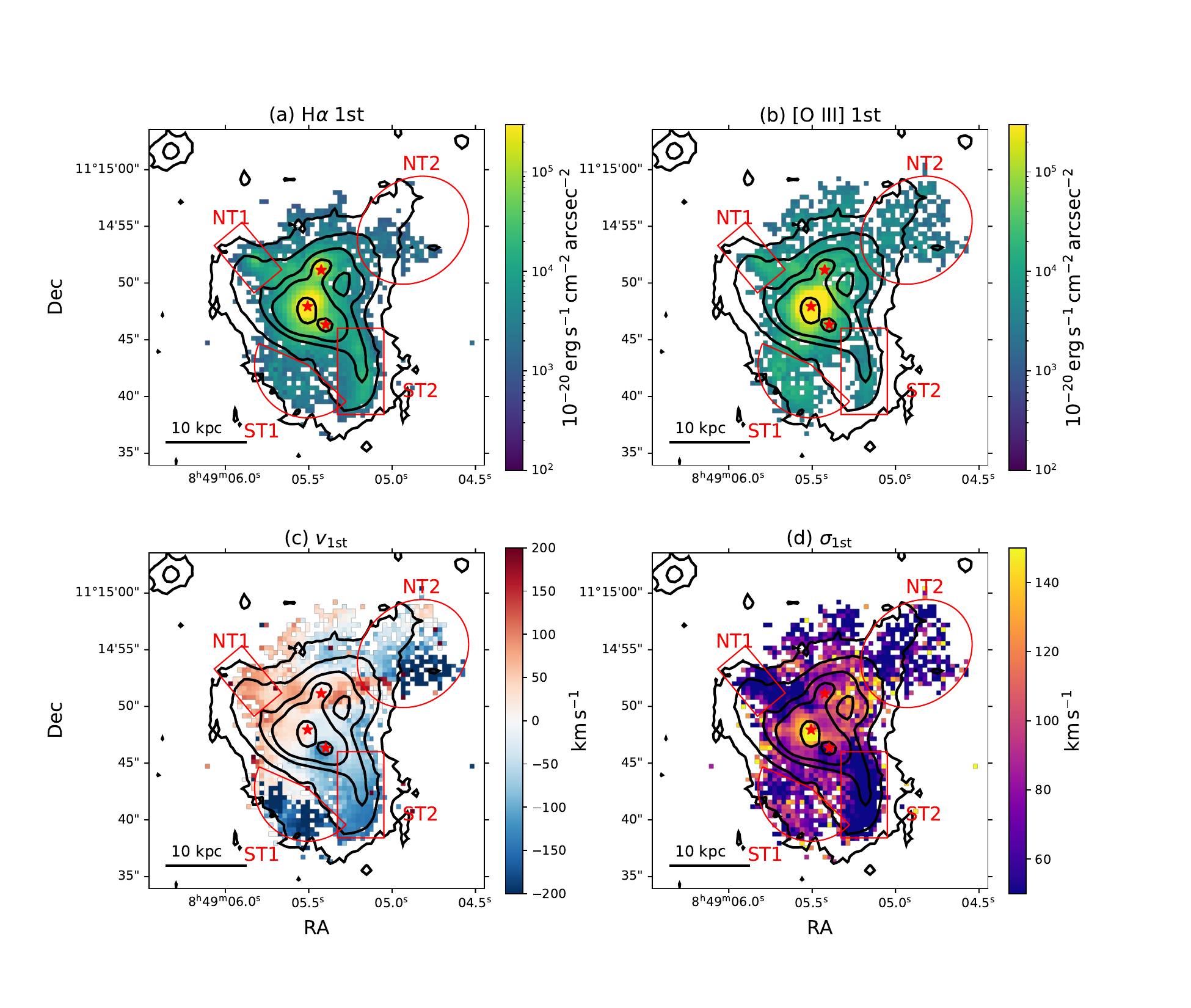}
\caption{Flux and kinematics maps of the first gas component. 
(a) and (b): Flux maps of the  H$\alpha$ and [O~{\sc{iii}}] emission lines, respectively.
Black contours are the stellar continuum emission from MUSE data (Figure~\ref{fig:hst_muse_spec}).
The regions NT1, NT2, ST1, and ST2 are shown in red.
(c) and (d): Velocity and velocity dispersion maps of the first emission line component.
Velocities are referenced to the first component of nucleus A in panel (c).
The locations of nuclei A, B, and C are marked with three red stars.
Only spaxels with S/N $\ge5$ and line flux $\ge 1\times10^{-18}\rm\,erg\,s^{-1}\,cm^{-2}$ were retained.
}
\label{fig:flux_kinematics_1st}
\end{figure*}

\begin{figure*}[ht!]
\includegraphics[width=1\textwidth,trim=0 30 0 20]{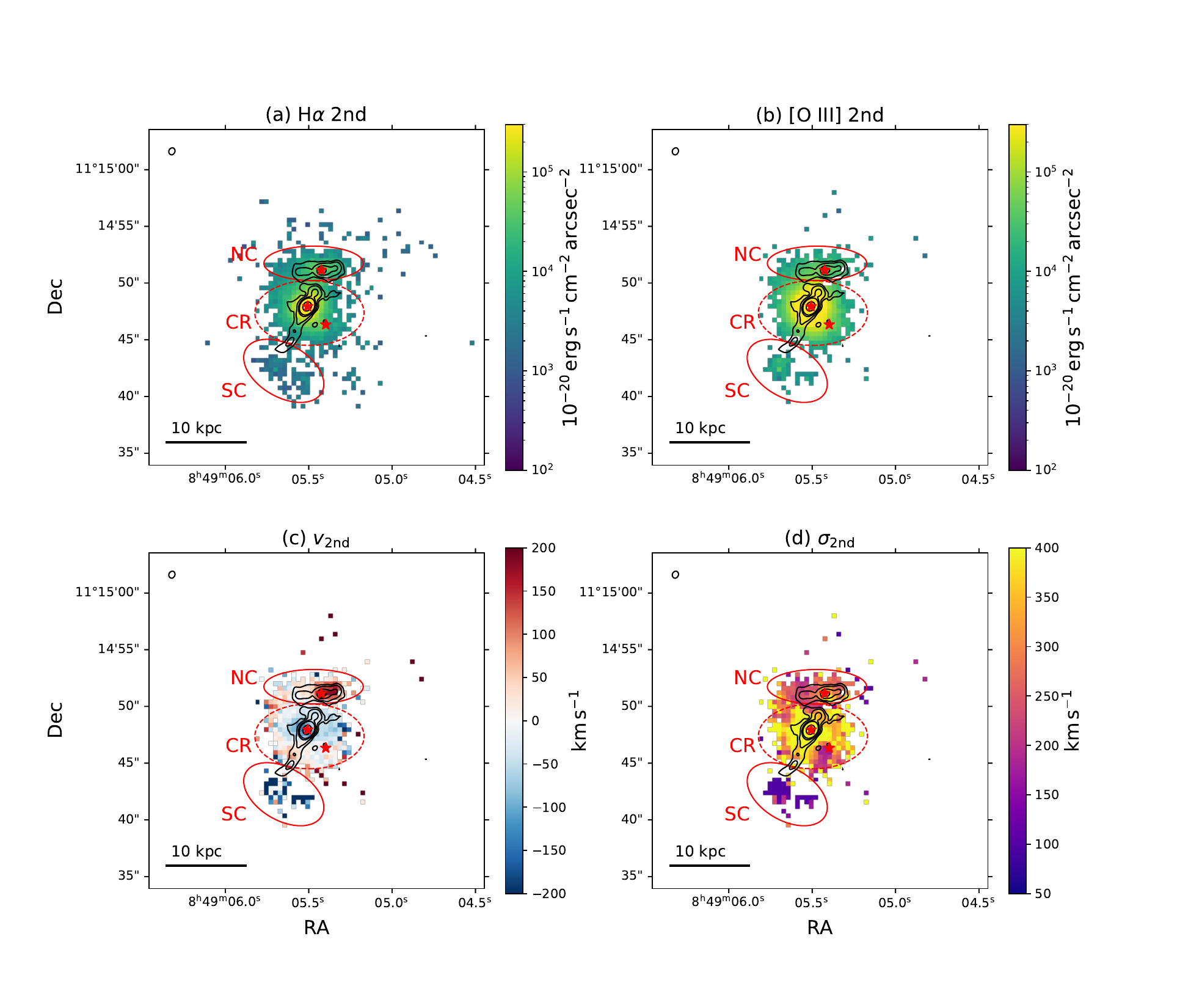}
\caption{Similar to Figure~\ref{fig:flux_kinematics_1st}, but for the second gas component. 
Black contours show the VLA 3 GHz emission from \cite{2022ApJ...934...89P}, with flux levels of 20, 80, 320, 640, and 5000 $\mu \rm Jy\,beam^{-1}$.
The regions NC and SC are marked with red ellipses, and the central region (region CR) is indicated by a red dashed ellipse.
}
\label{fig:flux_kinematics_2nd}
\end{figure*}

The first ionized-gas component traces rotation and tidal tails with low velocity dispersion, while the second traces outflows with higher velocity dispersion.
Here, we present the distribution and kinematics of each component.

Figure~\ref{fig:flux_kinematics_1st} shows the flux and kinematics maps of the first component of H$\alpha$ and [O~{\sc{iii}}] emission lines.
Regions NT1, NT2, ST1, and ST2 highlight the extended features associated with the first gas component.
These gas structures extend $>10$ kpc, and in some cases, even more than 15 kpc in projection (e.g. region NT2), away from nucleus A.
Compared to the continuum map from MUSE (black contours in Figure~\ref{fig:flux_kinematics_1st}), extended structures in regions NT1, NT2, and ST2 align well with tidal tails or diffuse stellar emission. 
The HST U-band image (Figure~\ref{fig:hst_muse_spec}) also show clear tidal tails in regions NT1 and ST2, which suggests that these features could originate from tidal stripping.
Region ST1 also exhibits diffuse stellar emission in the MUSE continuum map, but no clear tidal tail is visible in the HST images.
As the kinematics maps shown in Figure~\ref{fig:flux_kinematics_1st}, the extended gas features in regions NT1, NT2, ST1, and ST2 all exhibit low velocity dispersions, with median $\sigma_{\rm 1st}<70\rm\, km/s$.
The first component velocity ($v_{\rm 1st}$) in regions ST1 and ST2 is blueshifted, and $v_{\rm 1st}$ in region ST1 differs significantly from that around nuclei A and B.
$v_{\rm 1st}$ in region ST2 shows an decreasing trend from $-90$ to $-120\rm\,km\,s^{-1}$ from north to the south, and $v_{\rm 1st}$ in region ST1 also decrease from $-140$ to $-270\rm\,km\,s^{-1}$ from southwest to northeast. 
The findings suggest that the ionized gas in ST1 and ST2 might share a common origin; ST1 might be the downstream extension of the tidal tail observed in ST2.
Because the galaxy is undergoing a violent merger, alternative origins for the gas in ST1 cannot be excluded. 
Additional evidence is needed to pinpoint its true origin.
The velocity field around nucleus A is rotation-dominated, indicating the presence of a gaseous disk.

For the second component of the emission lines, as shown in Figure~\ref{fig:flux_kinematics_2nd}, extended structures are also found in three regions: region NC (northern cloud), region CR (central region), and region SC (southern cloud).
When compared to the VLA 3 GHz contours, the ionized gas emission in region NC is roughly consistent with the radio jet.
Region NC spans a projected length of $\sim 13\rm\,kpc$ from its eastern to its western end.
In Figure~\ref{fig:flux_kinematics_2nd}(c), $v_{\rm 2nd}$ in region NC shows an increasing gradient from east to west with $v_{\rm 2nd}\sim 40$--$180\rm\, km\,s^{-1}$.
The ionized gas in region NC also exhibits relatively high velocity dispersion (median $\sigma_{\rm 2nd}\sim250\rm\, km\,s^{-1}$).
Considering the $v_{\rm 2nd}$ of nucleus C is $\sim 150\rm\, km\,s^{-1}$, this velocity gradient suggests a bipolar outflow-like feature along the radio jet of nucleus C.

In region CR, the ionized gas displays brighter [O~{\sc{iii}}] emission compared to H$\alpha$, with its morphology resembling an ellipse or a box.
In the $v_{\rm 2nd}$ map, region CR exhibits blueshifted $v_{\rm 2nd}$ toward the north and redshifted $v_{\rm 2nd}$ toward the southeast, which is roughly aligned with the radio jet.
The velocity dispersion ($\sigma_{\rm 2nd}$) in region CR is large, with an mean $\sigma_{\rm 2nd}\sim 350 \rm\,km\,s^{-1}$.
These results suggest that the ionized gas (2nd component) in region CR could be originated from a bipolar outflow.
Low value of $v_{\rm 2nd}\sim 50 \rm\,km\,s^{-1}$ but large $\sigma_{\rm 2nd}\sim 350 \rm\,km\,s^{-1}$ in region CR imply the inclination angle between the outflow and the line-of-sight could be large.

In region SC, the second component reveals an isolated cloud of ionized gas.
Its blueshifted velocity ($v_{\rm 2nd}\sim -120\rm\,km\,s^{-1}$) and low dispersion ($\sigma_{\rm 2nd}\sim 100\rm\,km\,s^{-1}$) suggest that it is unlikely to be a nuclear outflow, and instead, it may be part of a tidal tail.

\begin{deluxetable}{cccc}
\tablenum{1}
\tablecaption{Mean kinematics of the regions.\label{tab:kinematics of regions}}
\tablewidth{0pt}
\tablehead{
\colhead{Region} & \colhead{gas component} & \colhead{$v$ (km/s)} & \colhead{$\sigma$ (km/s)}
}
\decimalcolnumbers
\startdata
NT1 & 1st & 75   & \textless{}50 \\
NT2 & 1st & -120 & \textless{}50 \\
ST1 & 1st & -170 & 65            \\
ST2 & 1st & -110 & \textless{}50 \\
NC  & 2nd & 80   & 250           \\
CR  & 2nd & -30  & 350           \\
SC  & 2nd & -120 & 100           \\
\enddata
\tablecomments{(1) Names of the regions. (2) The gas component from which the regions are derived. (3) The mean velocity ($v$) of each region referenced to the first component of nucleus A. (4) The mean velocity dispersion ($\sigma$) of each region. For regions NT1, NT2, and ST2, the velocity dispersion ($\sigma$) in most spaxels is unresolved and fixed at the lower limit of 50 km/s.}
\end{deluxetable}

\subsection{Emission-line diagnostic} \label{subsec:bpt-map}

\begin{figure*}[ht!]
\includegraphics[width=1\textwidth]{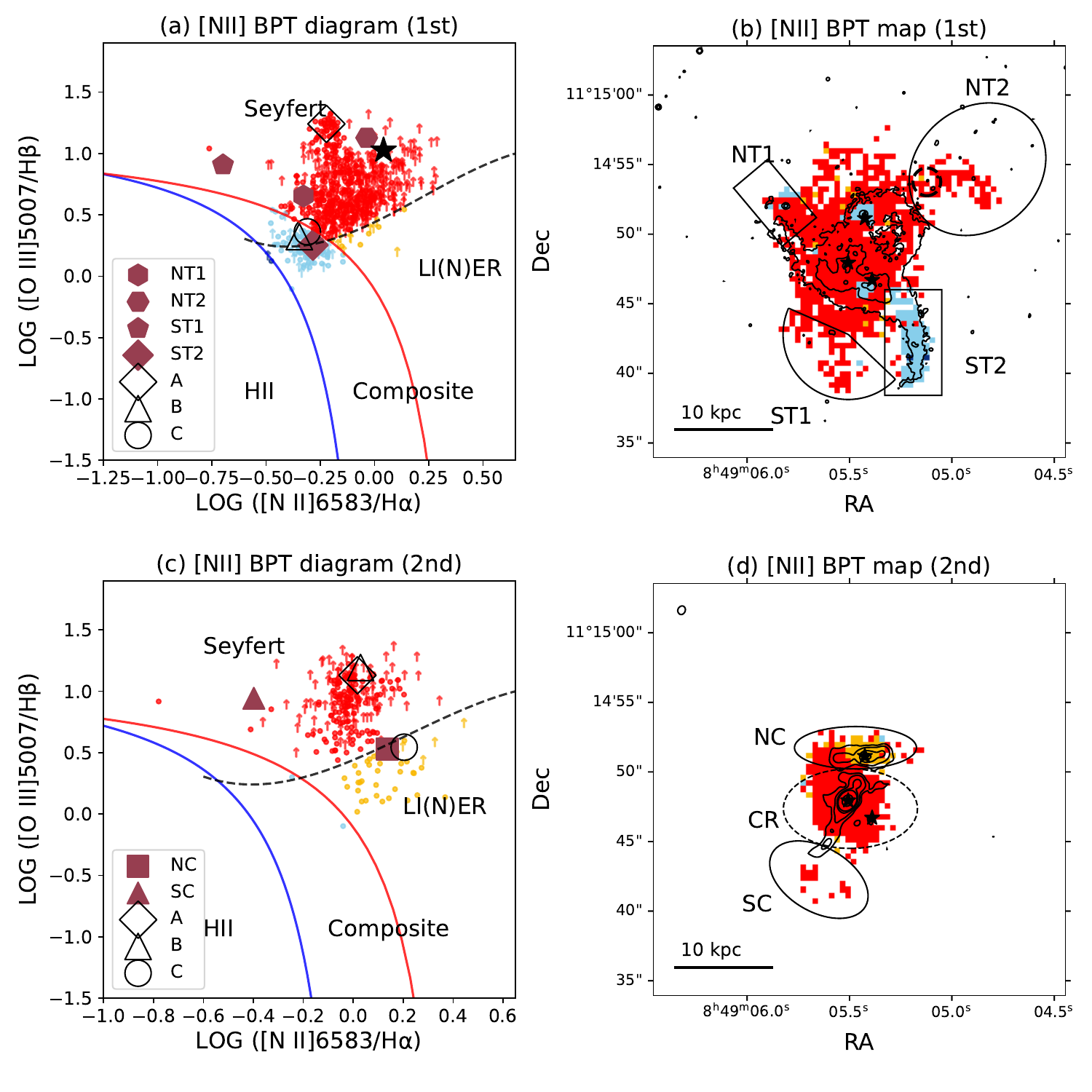}
\caption{BPT [N~{\sc{ii}}] diagnostic diagrams and maps.
Panels (a) and (b) show results for the first component, while panels (c) and (d) present results for the second component.
The blue and red solid lines are defined by \cite{2003MNRAS.346.1055K} and \cite{2001ApJ...556..121K}.
The black dashed line from \cite{2025arXiv250517843C} is used to separate the Seyfert and LINER regions.
In the BPT diagrams, three nuclei are marked with open markers based on measurements taken from a 1\arcsec-radius aperture centered on each nucleus.
The black star shows the integrated line ratios measured within the dashed circular in region NT2 (panel b).
Upper limits for log([O~{\sc{iii}}]/H$\beta$) are indicated with upward arrows.
}
\label{fig:bpt}
\end{figure*}

In Figure~\ref{fig:bpt}, the BPT [N~{\sc{ii}}] diagram and map are presented.
For spaxels where H$\beta$ is undetected, an upper limits for the H$\beta$ flux is set as $F_{\rm H\beta}=F_{\rm H\alpha}/3$.
For the first component of the ionized gas, most spaxels are classified as AGN photoionization in these two diagrams.
Spaxels in region ST2 and some spaxels around nuclei B and C are identified as composite regions in the BPT map.
Nuclei B and C are located at the composite region, which might seem at odds with \cite{2019ApJ...887...90L} and \cite{2019ApJ...883..167P}. 
In reality, line ratios here differ from theirs by only a small margin, so the overall conclusions remain consistent.
The BPT results of Nuclei B and C could be diluted by the host galaxy emission, whereas X-ray observations still confirm that both nuclei are AGNs \citep{2019ApJ...887...90L}.

For the second component of the ionized gas, all spaxels in regions CR and SC are dominated by AGN photoionization.
In Figure~\ref{fig:bpt}(d), spaxels around region NC is primarily classified as LI(N)ER, indicating that the ionized gas in this area may be shocked by the radio emission or dominated by emission from an AGN with a low accretion rate.

\section{Discussion} \label{sec:discussion}

\subsection{Energetics and origins of the ionized gas outflows} \label{subsec:origin}

As shown in Sect~\ref{sec:result}, the second component of the ionized gas in the region CR and region NC could be the ionized gas outflows, while the first component is dominated by the rotational disk and tidal tails.

The ionized gas mass and outflow rates can be estimated using $\rm H\beta$ luminosity of 2nd component following previous works \citep[e.g.][]{2006agna.book.....O,2022ApJ...933..110X}.
The outflow velocity could be estimated with $v_{\rm out} = v_{\rm 2nd} + 2\sigma_{\rm 2nd}$ \citep[e.g.][]{2017A&A...601A.143F}.
For the region CR and region NC, we adopt outflow velocity $v_{\rm out,A} = v_{\rm 2nd,A} + 2\sigma_{\rm 2nd,A} = 800\,\rm km\,s^{-1}$ and $v_{\rm out,C} = (v_{\rm 2nd,C}-v_{\rm 1st,C}) + 2\sigma_{\rm 2nd,C} = 670\,\rm km\,s^{-1}$ (relative to nucleus C), respectively.
Adopting the electron density $n_{\rm e,A}= 1500 \rm\, cm^{-3}$ and the outflow size $R_{\rm out,A} = 5.3 \rm\, kpc$, we obtain the mass outflow rate $\dot{M}_{\rm out,A} = 14.8 \rm\, M_{\astrosun}\, yr^{-1}$ and kinetic power $\dot{E}_{\rm out,A} = 3.0\times10^{42}\rm\, erg\, s^{-1}$.
Here, the electron density is calculated from the [S~{\sc{ii}}]$\lambda\lambda 6716,6731$ (2nd gas component) line ratio.
Considering the luminosity of the second component of [O~{\sc{iii}}] around nucleus B ($r<1\arcsec$) is $<5\%$ of the luminosity around nucleus A, we neglect the contribution of nucleus B to the outflowing gas in the central region.
Similarly, for region NC, we derive $\dot{M}_{\rm out,C} = 0.1 \rm\, M_{\astrosun}\, yr^{-1}$ and $\dot{E}_{\rm out,C} = 2.0\times10^{40}\rm\, erg\, s^{-1}$, adopting $n_{\rm e,C}= 1200 \rm\, cm^{-3}$, $R_{\rm out,C} = 5.9 \rm\, kpc$, and $v_{\rm out,C} = 690\,\rm km\,s^{-1}$.

In recent years, many studies have highlighted that estimates of mass and energy outflow rates can vary by up to an order of magnitude, depending on the assumptions and parameters used, such as ionized gas mass, velocities, and densities \citep[e.g.][]{2017A&A...601A.143F,2021MNRAS.504.3890D}.
As discussed in \cite{2021MNRAS.504.3890D} and \cite{2017A&A...601A.143F}, different definitions of outflow velocity can lead to variations in the estimated mass outflow by several factors.
When accounting for uncertainties in various parameters (e.g., outflow velocity, size, geometry, etc.), \cite{2024A&A...691A..15D} estimated that the outflow rate measurements would exhibit a typical uncertainty factor $\sim5$ at 1$\sigma$ confidence level.
Electron density ($n_{\rm e}$) is another critical parameter. 
It is typically estimated using the [S~{\sc{ii}}]$\lambda 6716/$[S~{\sc{ii}}]$\lambda 6731$ line ratio, assuming an electron temperature of $T_{e} = 10^{4}\rm\,K$.
Reported values for electron densities derived from the [S~{\sc{ii}}] doublet range from $\sim 100$ to $2000\rm\, cm^{-3}$ \citep[e.g.][]{2019ApJ...875...21F,2021MNRAS.504.3890D}.
For nuclei A and C, we estimate $n_{\rm e,A}= 1500 \rm\, cm^{-3}$ and $n_{\rm e,C}= 1200 \rm\, cm^{-3}$, consistent with values found in the literature.
However, the reliability of [S~{\sc ii}] line ratios as electron density diagnostics has been questioned by several authors \citep[e.g.][]{2018NatAs...2..198H,2020MNRAS.498.4150D}.
Alternative tracers—such as [Ar~{\sc{iv}}]$\lambda\lambda 4711,4740$, [S~{\sc{ii}}]$\lambda\lambda 4069,4076$, and [O~{\sc{ii}}]$\lambda\lambda 7320,7331$—have been proposed to better constrain electron densities in AGNs \citep[e.g.][]{2014A&A...561A..10P,2011MNRAS.410.1527H,2020MNRAS.498.4150D}.
These diagnostics often yield electron densities several to ten times higher than those derived from the [S~{\sc{ii}}]$\lambda\lambda 6716,6731$ lines ratio.
Unfortunately, these alternative lines are either too faint or fall outside the wavelength coverage of our MUSE data.
While our estimates of outflow rates may carry significant uncertainties, the first-order values still provide valuable insight into the impact of AGN-driven outflows.

Ionized outflows can be launched by AGN radiation, starbursts, or radio jets.

Radio jet-driven warm ionized gas outflows have been observed in some galaxies \citep[e.g.][]{2019MNRAS.485.2710J,2021A&A...648A..17V}.
In region CR, the velocity dispersion of the outflow is enhanced perpendicular to the radio jet, which is consistent with the jet driving scenario \citep[e.g.][]{2021A&A...648A..17V}.
Additionally, the velocity gradient in region CR is also along the radio jet of nucleus A (Fig~\ref{fig:flux_kinematics_2nd}).
Energetics support the same picture.
Using the same 3 GHz data and method as \cite{2022ApJ...934...89P} (assuming an inclination angle of $45^{\circ}$), the jet energy is $E_{\rm jet,A}\sim 4.9\times10^{57}\rm\, erg$.
Adopting the dynamical timescale $t_{\rm out,A} = R_{\rm out,A}/v_{\rm out,A} \simeq 6.5 \rm\,Myr$, the ionized outflow of nucleus A carries $E_{\rm out,A} \simeq 6.1 \times10^{56}\rm\, erg\sim 12\%\times E_{\rm jet,A}$, which is consistent with the $\leq30\%$ coupling efficiency predicted by simulations \citep[e.g.][]{2011ApJ...728...29W,2016MNRAS.461..967M}.
Star formation cannot supply this power.
An SFR of $\rm SFR_{\rm A} = 8-13.16\,M_{\astrosun}\,yr^{-1}$ for nucleus A \citep[][]{2019ApJ...887...90L,2019ApJ...883..167P} yields only $\dot{E}_{\rm SN,A}\sim 1.7-2.8\times10^{41}\rm\, erg\,s^{-1}$ \citep[following][]{2017A&A...601A.143F}, $\sim 10$ times weaker than $\dot{E}_{\rm out,A}$.
If the ionized gas outflow is driven by the AGN radiation, the required AGN bolometric luminosity would be $\sim 10^{45}\rm\, erg\, s^{-1}$ according to the scaling relation from \cite{2017A&A...601A.143F}.
From the 2–-10 keV luminosity \citep[$L_{\rm X,A} = 1.3\times 10^{42}\rm\, erg\, s^{-1}$][]{2019ApJ...887...90L} and the bolometric correction from \cite{2020A&A...636A..73D}, the current bolometric luminosity of nucleus A is estimated to be $L_{\rm bol,A,X} = 2.1_{-1}^{+2}\times 10^{43}\rm\, erg\, s^{-1}$, accounting for a bolometric correction uncertainty of $\sim 0.3\rm\,dex$.
Adopting the SMBH mass of $M_{\rm BH,A} = 10^{7.9} \,\rm M_{\odot}$ estimated by \cite{2019ApJ...887...90L}, the Eddington luminosity of nucleus A is $L_{\rm Edd,A}\sim 1.1\times 10^{46}\rm\, erg\, s^{-1}$.
Therefore, the Eddington ratio for nucleus A is $L_{\rm bol,A,X}/L_{\rm Edd,A} \sim 0.002_{-0.001}^{+0.002}$.
The current AGN is therefore too weak to power the observed outflow, although a luminous phase $\sim 6.5 \rm\,Myr$ ago cannot be ruled out. 
Thus, the ionized outflow from nucleus A is most plausibly powered by its radio jet.

The ionized gas outflow in region NC shows a velocity gradient along the radio jet, suggesting that the outflow is associated with the radio jet of nucleus C.
\cite{2022ApJ...934...89P} estimated the total energy of the jet of nucleus C to be $E_{\rm jet,C}=5.0\times10^{55}\rm\, erg$.
The ionized outflow carries $E_{\rm out,C} \simeq 5.2 \times10^{54}\rm\, erg$—about $10\%$ of $E_{\rm jet,C}$ of the jet energy—matching simulation expectations.
With a SFR of only $\rm SFR_{\rm C} = 0.1\,M_{\astrosun}\,yr^{-1}$ for nucleus C \citep[][]{2019ApJ...887...90L}, a starburst wind would supply far less kinetic power than we measure in the ionized outflow. 
Star formation alone therefore cannot drive the observed gas outflow.

\subsection{Episodic AGN activities and feedback} \label{subsec:episodic feedback}

\begin{figure*}[ht!]
\includegraphics[width=1\textwidth]{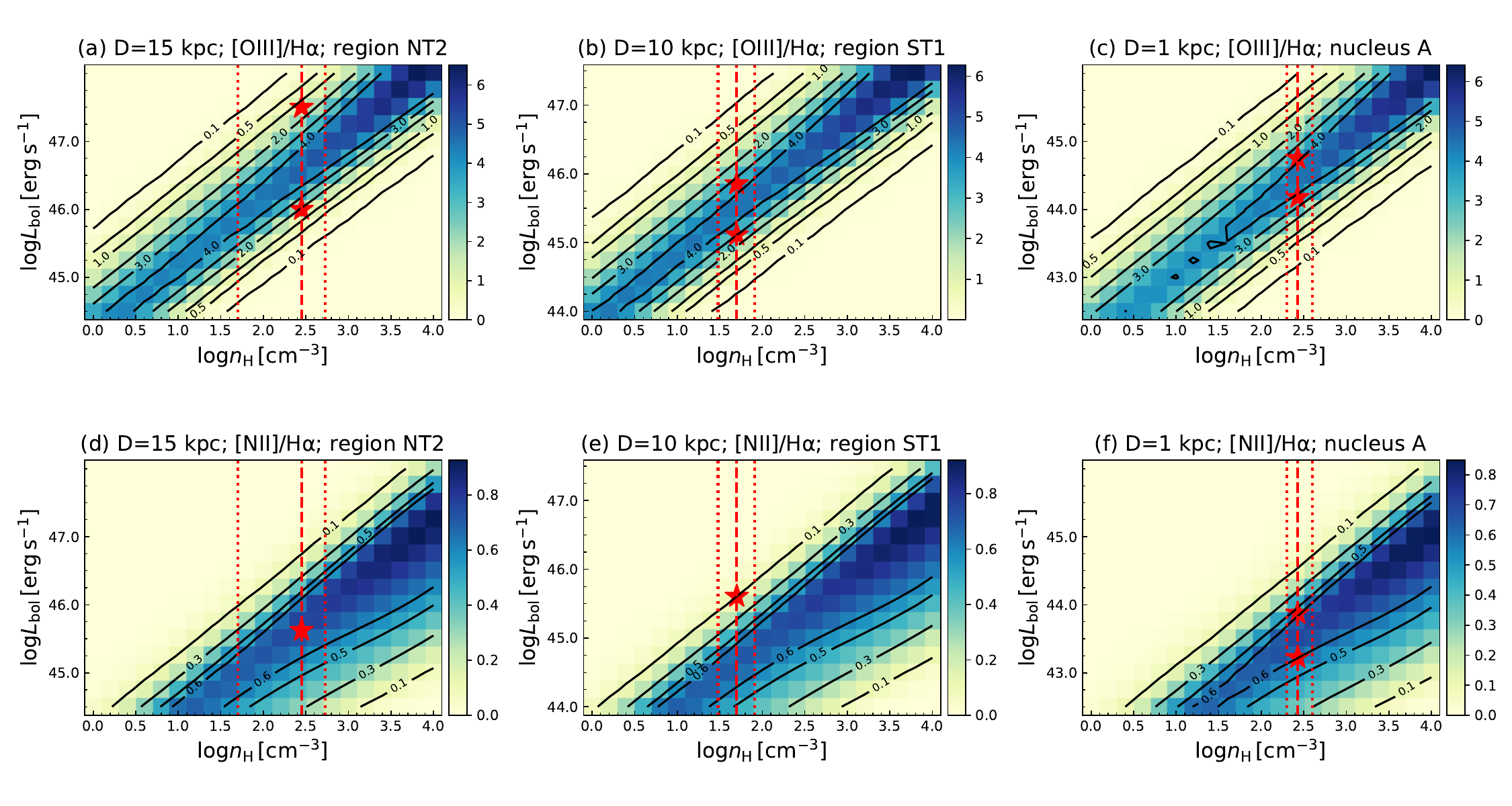}
\caption{(a): [O~{\sc{iii}}]/H$\alpha$ ratios derived from photoionization models of CLOUDY, assuming a typical AGN SED \citep[][]{2012MNRAS.420.1825J} and the cloud is 15 kpc away from the AGN.
Red stars are determined by the intersection of the electron density of $280\rm\,cm^{-3}$ (red dashed line) and the [O~{\sc{iii}}]/H$\alpha$ ratio of $\sim 1.6$ estimated from the dashed circle region in region NT2 (see Figure~\ref{fig:bpt}).
The dotted lines indicate the 1$\sigma$ uncertainty range of the electron density.
(b) and (c): Similar to (a), but assuming a cloud with a distance of 10 kpc (region ST1) and 1 kpc away from the AGN, respectively.
The electron densities of $50\rm\,cm^{-3}$ and $270\rm\,cm^{-3}$, and [O~{\sc{iii}}]/H$\alpha$ of $\sim 3.2$ and $\sim 3.75$ are derived from region ST1 and nucleus A (aperture radius of 1\arcsec), respectively.
(d), (e), and (f): Similar to (a), (b), and (c), but for [N~{\sc{ii}}]/H$\alpha$ ratios.
}
\label{fig:cloudy}
\end{figure*}

Observations show that some AGNs could fade by orders of magnitude within $\sim 1$--$10\times10^{4}\rm\, yr$ \citep[e.g.][]{2012MNRAS.420..878K,2017ApJ...835..256K,2022ApJ...938...75P}.
J0849+1114 is a merging system, which could lead to more rapid and violent variations in central AGN activity.

Extended tidal tails ($>10\rm\, kpc$) in regions NT1, NT2, and ST1 are dominated by AGN-photoionized (Figure~\ref{fig:bpt}), which suggest a luminous AGN in J0849+1114 to ionize these gas.
Using CLOUDY \citep[v23.01, last described by][]{2023RMxAA..59..327C,2023RNAAS...7..246G} and an AGN SED with a median Eddington ratio ($L_{\rm AGN}/L_{\rm Edd}=0.76$) from \cite{2012MNRAS.420.1825J}, we reproduced the observed [O~{\sc{iii}}]/H$\alpha$ and [N~{\sc{ii}}]/H$\alpha$ ratios in different regions (Figure~\ref{fig:cloudy}).
The electron density is estimated from the the [S~{\sc{ii}}]$\lambda\lambda 6716,6731$ line ratio (1st gas component).
the dashed circle region within NT2, the electron density is $280_{-230}^{+270}\rm\,cm^{-3}$, which is relatively high, though with significant uncertainty.
When the electron density is estimated from the integrated spectrum of the entire NT2 region, the density is found to be lower than $10\rm\,cm^{-3}$. This suggests that most spaxels in region NT2 have low electron densities, with only the dashed circle region, closer to the galaxy's main body, exhibiting a relatively higher density.
To ionize the dashed circle region in region NT2 ($\sim 15\rm\, kpc$ from nucleus A in projection, see Figure~\ref{fig:bpt}), it requires a bolometric luminosity of $L_{\rm A,bol} \sim 1.5$--$6\times10^{45}\rm\, erg\,s^{-1}$ ($\sim 0.1$--$0.5\,L_{\rm Edd,A}$), considering the electron density uncertainty. 
Figure~\ref{fig:cloudy}(a) suggests that the bolometric luminosity should be $L_{\rm A,bol} \sim 5\times10^{45}\rm\, erg\,s^{-1}$ or even larger than $10^{47}\rm\, erg\,s^{-1}$.
However, when combined with the results from Figure~\ref{fig:cloudy}(d), the bolometric luminosity should be lower than $10^{46}\rm\, erg\,s^{-1}$.
The derived Eddington ratio, $L_{\rm A,bol}/L_{\rm Edd,A}\sim 0.1$--$0.5$, is somewhat lower than that of the input AGN SED, which has $L_{\rm AGN}/L_{\rm Edd}=0.76$.
If an AGN SED with a lower Eddington ratio is adopted, the derived AGN luminosity would be slightly higher than the current estimate of $L_{\rm A,bol}$.
Therefore, the results based on the input AGN SED with a median Eddington ratio may slightly underestimate the true AGN luminosity, but this does not compromise the validity of our conclusions.

Similarly, for region ST1 ($\sim 10\rm\, kpc$ from nucleus A in projection), the required AGN luminosity is $L_{\rm A,bol} \sim 1$--$3\times10^{45}\rm\, erg\,s^{-1}$.
In contrast, the ionization of circumnuclear gas ($\sim 1\rm\, kpc$ from nucleus A) only requires $L_{\rm A,bol} \sim 0.8$--$1.5\times10^{44}\rm\, erg\,s^{-1}$. 
The current bolometric luminosity estimated from hard X-ray is $L_{\rm bol,A,X} = 2.1_{-1}^{+2}\times 10^{43}\rm\, erg\, s^{-1}$, which is even lower.
These comparisons suggest that nucleus A was 20--100 times more luminous $\sim 3$--$5\times10^{4}\rm\, yr$ ago (light travel time from nucleus A to region ST1 and NT2).
Because the [O~{\sc{iii}}] recombination timescale is $<100\rm\,yr$ for $n_{\rm e}\sim 100\rm\,cm^{-3}$, the ionized gas in tidal tails must have been sustained by a long-lived ionization source until several $10^{4}\rm\, yr$ ago.

We focus on nucleus A because it hosts the most massive SMBH, is currently the brightest of the three nuclei, and lies at the system’s center, making it the most plausible source of the extended ionization.
The possible contributions of nuclei B and C cannot be ruled out.
However, their smaller black-hole masses \citep[][]{2019ApJ...887...90L} may require Eddington ratios higher than that of nucleus A if they were the main sources ionizing the extended gas.

The mass loading factor ($\dot{M}_{\rm out}/\rm SFR$) is commonly used to quantify the impact of outflows on the host galaxy.
Based on the SFR of nucleus A, $\rm SFR_{\rm A} = 8\,M_{\astrosun}\,yr^{-1}$ \citep[][]{2019ApJ...887...90L}, and the mass outflow rate, $\dot{M}_{\rm out,A} = 14.8 \rm\, M_{\astrosun}\, yr^{-1}$, the mass loading factor is calculated as $\dot{M}_{\rm out,A}/\rm SFR_{\rm A} \simeq 1.8$.
However, the mass outflow rate could vary by several times (even an order of magnitude) according to the discussion in Sect~\ref{subsec:origin}.
If we consider the typical 1$\sigma$ uncertainty estimated by \cite{2024A&A...691A..15D}, which corresponds to a factor of $\sim 5$, the mass loading factor for nucleus A would be $\dot{M}_{\rm out,A}/\rm SFR_{\rm A} \simeq 1.8_{-1.4}^{+7.4}$.
This indicates that the ionized gas outflow could deplete more gas than the star formation in nucleus A, although the lower limit of the mass loading factor is below 1.
Similarly, for nucleus C, the mass loading factor is only $\dot{M}_{\rm out,C}/\rm SFR_{\rm C} \simeq 0.05_{-0.04}^{+0.2}$ \citep[$\rm SFR_{C}=2\,M_{\astrosun}\, yr^{-1}$ from][]{2019ApJ...887...90L}, suggesting that the ionized gas outflow in nucleus C plays a less significant role in depleting gas compared to SF of nucleus C.

\cite{2022ApJ...934...89P} reported two radio jets from nucleus A: an inner jet 0.\arcsec6 long at 15.0 GHz with a dynamical age of $\sim3\times 10^{4}\rm\, yr$ and an outer jet 5.\arcsec5 long at 3 GHz with a dynamical timescale of $\sim 1.5\times 10^{5}\rm\, yr$ (Figure~\ref{fig:flux_kinematics_2nd}).
Combining with our results, all these features may reveal episodic AGN activities.
The outer jet probably launched during previous active phase $\sim1.5\times 10^{5}\rm\, yr$ ago, and the ionized gas outflow could be driven by this outer jet.
A later high-accretion burst ionized the tidal tails.
The inner jet appeared $\sim 3\times 10^{4}\rm\, yr$ ago as the AGN began to fade, and the nucleus has remained in a low-accretion state ever since.
The outflow and jet of nucleus A can expel and/or heat circumnuclear gas and starve or regulate the SMBH.
Previous X-ray surveys of AGN pairs and galaxy pairs have found evidence for the self-regulation in late stage mergers \citep[][]{2020ApJ...900...79H,2023ApJ...949...49H}.
\cite{2020ApJ...900...79H} found that, in AGN pairs with projected separations $r_{\rm p}\leq 10\rm\,kpc$, the mean X-ray luminosity drops as the nuclei draw closer.
Further, \cite{2023ApJ...949...49H} confirmed this trend in a larger sample of close galaxy pairs.
They suggested that most SMBHs in these systems are currently accreting at a low rate because negative AGN feedback has cleared out the circumnuclear gas.
The outflow, possibly driven by a jet, and the low current accretion rate we infer for J0849+1114 fit naturally into this picture.

\section{Conclusions} \label{sec:summary}

In this work, we study the spatially resolved ionized gas in the triple AGN candidate J0849+1114 using VLT/MUSE data.
To identify the ionized gas outflows, two-component Gaussian fitting was used to analysis the datacube.
For the first Gaussian component of the ionized gas, several extended structures are detected in J0849+1114, which primarily originate from tidal tails based on their morphology and kinematics.
The second component reveals two ionized gas outflows around nuclei A and C, both likely powered by their radio jets.
Based on the calculation of CLOUDY, the nucleus A was likely a luminous quasar with $L_{\rm A,bol} \sim 0.1$--$0.5\,L_{\rm Edd,A}$ at least $\sim 3$--$5\times10^{4}\rm\, yr$ ago.
Combining the dynamical timescales of the radio jets, we might find evidence of episodic activities of nucleus A during the past $1\times10^{5}\rm\, yr$: 
(i) $1.5\times10^{5}\rm\, yr$ ago an active phase launched the outer jet, which could drive the ionized gas outflow;
(ii) a subsequent luminous burst ionized the tidal tails;
(iii) $\sim 3\times 10^{4}\rm\, yr$ ago the inner jet appeared as the AGN began to fade, and the AGN has remained in a low-accretion state ever since.
Our results show that, in J0849+1114, AGN feedback can rapidly suppress central SMBH accretion and impact the host on kpc scales.

\begin{acknowledgments}

We thank Xin Liu the helpful comments and suggestions that strengthened this work.
X.X. acknowledges the NSFC grant 12403018, the China Postdoctoral Science Foundation (No. 2023M741639), and the Jiangsu Funding Program for Excellent Postdoctoral Talent (No. 2024ZB249).
M.H. acknowledges support from the NSFC grant 12203001.
Z.L. acknowledges the National Natural Science Foundation of China (grant 12225302).
J.W. acknowledges the NSFC grants 12333002, 12033004, 12221003, and China Manned Space Project with No. CMS-CSST-2025-A07 and CMS-CSST-2025-A10.
Based on observations collected at the European Southern Observatory under ESO program 108.21ZY.012.

\end{acknowledgments}

%

\vspace{5mm}
\facilities{VLT:MUSE}


\software{astropy \citep{2013A&A...558A..33A}, EsoReflex \citep{2013A&A...559A..96F}, PPXF \citep{2004PASP..116..138C}, DS9 \citep{2003ASPC..295..489J}, CLOUDY \citep{2023RMxAA..59..327C,2023RNAAS...7..246G}
          }



\appendix

\section{Spectra of the Three Nuclei and Regions NC and SC} \label{sec:spectra_example}

\begin{figure*}[ht!]
\includegraphics[width=0.85\textwidth]{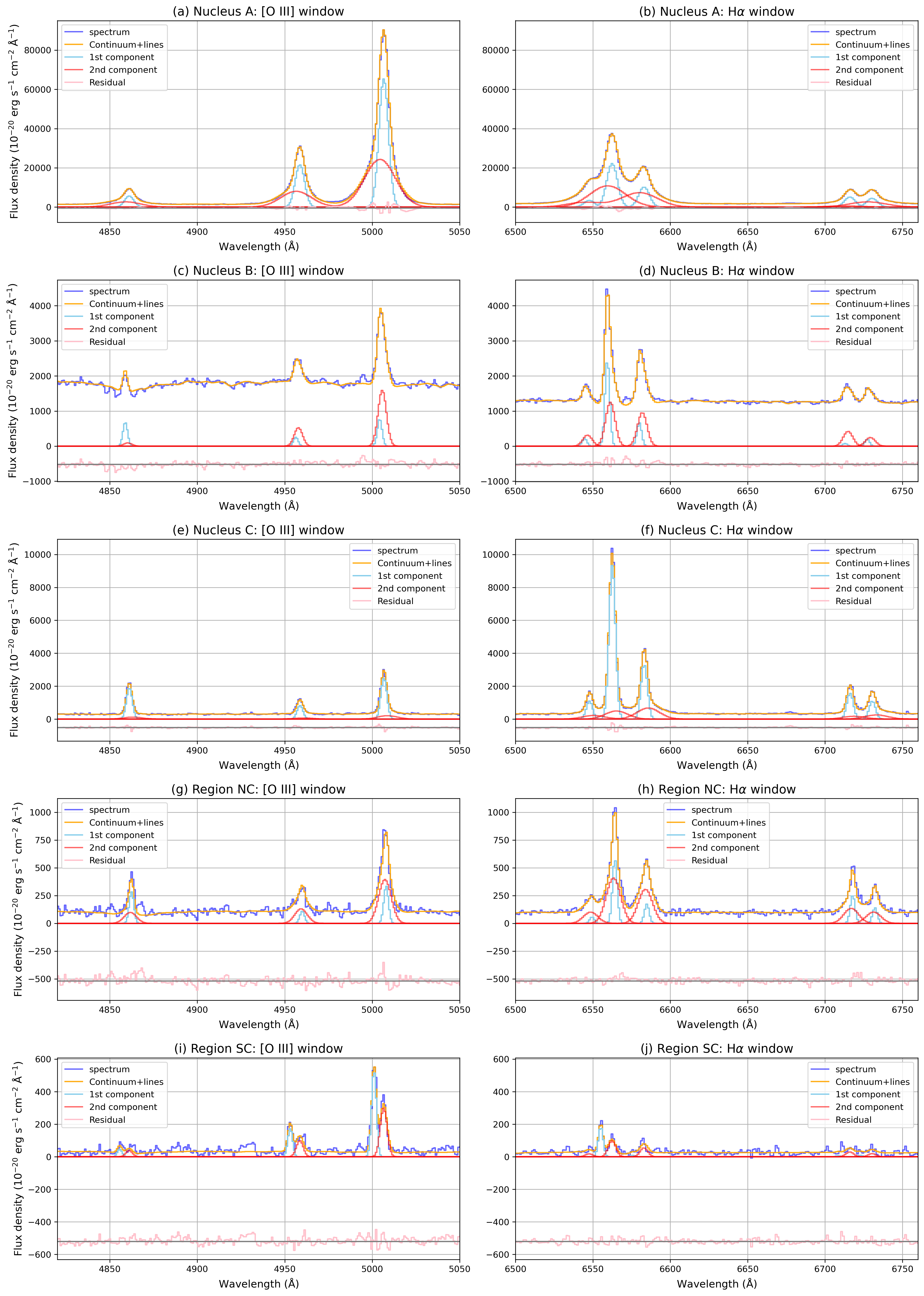}
\caption{Zoomed-in spectra for the rest-frame wavelength ranges $4820$--$5050\rm\,\AA$ (left) and $6500$--$6760\rm\,\AA$ (right).
(a) and (b): Spectrum of nucleus A.
(c), (d), (e), and (f): same as (a) and (b), but for nuclei B and C, respectively.
(g) and (h): Spectrum extracted from the central spaxel of east radio lobe in region NC.
(i) and (j): Spectrum extracted from the central spaxel of region SC.
The observed data is shown in blue, while the total model (continuum plus emission lines) is shown in orange.
The two Gaussian components are represented by light blue and red lines, respectively.
The residuals are displayed in pink.
}
\label{fig:spectra}
\end{figure*}

Figure \ref{fig:spectra} presents the spectra of the three nuclei and of regions NC and SC. 
In each case the main emission lines are well described by two Gaussian components. 
Including a third Gaussian component cannot significantly improve the fit anywhere in the FoV.
\cite{2019ApJ...875..117P} likewise reported double-component emission lines in all three nuclei.

\bibliography{sample631}{}
\bibliographystyle{aasjournal}


\end{CJK*}
\end{document}